# Pressure-control of non-ferroelastic ferroelectric domains in ErMnO$_3$

O. W. Sandvik[1], A. M. Müller[2], H.W. Ånes[1], M. Zahn[1,3], J. He[1], M. Fiebig[2], Th. Lottermoser[2], T. Rojac[4], D. Meier[1,*], and J. Schultheiß[1,3,*]

[1] Department of Materials Science and Engineering, Norwegian University of Science and Technology (NTNU), 7034 Trondheim, Norway
[2] Department of Materials, ETH Zurich, 8092 Zurich, Switzerland
[3] Experimental Physics V, University of Augsburg, 86159 Augsburg, Germany
[4] Electronic Ceramics Department, Jožef Stefan Institute, 1000 Ljubljana, Slovenia

*corresponding authors: dennis.meier@ntnu.no; jan.schultheiss@uni-a.de

**Mechanical pressure controls the structural, electric, and magnetic order in solid state systems, allowing to tailor and improve their physical properties. A well-established example is ferroelastic ferroelectrics, where the coupling between pressure and the primary symmetry breaking order parameter enables hysteretic switching of the strain state and ferroelectric domain engineering. Here, we study the pressure-driven response in a non-ferroelastic ferroelectric, ErMnO$_3$, where the classical stress-strain coupling is absent, and the domain formation is governed by creation-annihilation processes of topological defects. By annealing ErMnO$_3$ polycrystals under variable pressures in the MPa-regime, we transform non-ferroelastic vortex-like domains into stripe-like domains. The width of the stripe-like domains is determined by the applied pressure as we confirm by three-dimensional phase field simulations, showing that pressure leads to highly oriented layer-like periodic domains. Our work demonstrates the possibility to utilize mechanical pressure for domain engineering in non-ferroelastic ferroelectrics, providing a processing-accessible lever to control their dielectric, electromechanical, and piezoelectric response.**

## 1. Introduction

The functionality of ferroelectrics is intimately coupled to their domain structure,[1] which allows for tuning the dielectric, piezoelectric, and electromechanical properties by introducing, for example, morphotropic phase boundaries,[2] critical points,[3] or defect-complexes[4]. Another well-established and very versatile approach for property engineering are elastic strains, which have been applied to enhance the spontaneous polarization in BiFeO$_3$ thin films,[5] stabilize ferroelectric order in SrTiO$_3$[6], and tailor the performance of BaTiO$_3$ via extended defects[7] or internal microstructural effects[8].

In addition, the application of an external mechanical pressure has substantial impact on the formation of domains, co-determining their size, shape, and stability and, hence, the macroscopic ferroelectric responses[9-11]. Most existing concepts for ferroelectric domain engineering via elastic strain focus on the family of perovskite oxides, exploiting strain-induced ferroelastic domain walls that form in addition to the ferroelectric ones to release the strain.[12, 13] More recently, strain-driven domain engineering in non-ferroelastic ferroelectrics, i.e., ferroelectrics that do not host ferroelastic domain walls, has drawn attention. In particular, improper ferroelectric hexagonal manganite (RMnO$_3$, R = Sc, Y, In, and Dy-Lu),[14, 15] which is isostructural to hexagonal ferrite, indate, and gallate,[16, 17] has been studied. Ferroelectricity in RMnO$_3$ arises as a secondary effect, driven by a trimerizing lattice distortion, which represents the primary symmetry breaking order parameter.[18] As a consequence, the ferroelectric domain structure of RMnO$_3$ exhibits characteristic sixfold meeting points of alternating ±P domains (P denotes the local polarization),[19] forming topologically protected vortex-/anti-vortex pairs. Related to the improper nature, RMnO$_3$ provides a large variety of unusual physical phenomena, ranging from charged domain walls with unique electronic properties[19-21] to non-conventional domain-scaling behavior[22-24]. Importantly, the structural breaking order parameter in RMnO$_3$ is coelastic rather than ferroelastic, and the individual ferroelectric domain walls are not expected to move in response to an externally applied stress.[25] Interestingly, domain structure engineering can be anyway realized via a strain gradient. For the hexagonal manganites, an elastic strain gradient creates a pulling force on the vortex-/anti-vortex pair, resulting in a transformation of the isotropic vortex-like domains into stripe-like patterns[26] and an inversion of the domain scaling behavior with respect to grain size compared to classical perovskite systems[27]. These findings reflect completely different domain physics beyond what is known from ferroelastic ferroelectrics, representing a largely unexplored playground for the engineering of ferroelectric domains and polar nanostructures.

Here, we study the impact of uniaxial pressure up to the MPa regime on the domain formation in ErMnO$_3$ polycrystals during high-temperature treatment through the Curie temperature. Our systematic analysis reveals a coherent response of the uniaxial ferroelectric grains, leading to a decrease in domain size with increasing pressure. Interestingly, despite the random crystallographic orientation of the grains with respect to the direction of the applied pressure, we observe that all grains transform from an isotropic vortex-like domain structure to stripe-like domains as a function of the pressure, with the domain walls orienting parallel to the polar axis of each grain. This behavior is fundamentally different from the pressure-response of non-ferroelastic ferroelectrics, where the domain engineering is not possible via an applied mechanical pressure.

## 2. Results and Discussion

### 2.1. Domain morphology after high-pressure annealing

To study the impact of pressure on the domain formation in non-ferroelastic ferroelectrics, we utilize the hexagonal manganite system ErMnO$_3$. Phase-pure ErMnO$_3$ polycrystals are synthesized via a solid-state approach at a temperature of 1400°C, which allows for obtaining dense samples with a cylindrical geometry (diameter: ~7.6 mm, height: ~5.1 mm, see methods and ref. [27] for details). After synthesis, our polycrystals exhibit the typical *R*MnO$_3$ domain structure as discussed in detail in ref. [27]. To study the impact of uniaxial pressure on the domain structure, we anneal our samples above the Curie temperature ($T_c$ ~ 1156°C[28]) at 1220°C and apply a constant uniaxial mechanical pressure along the long axis of the cylindrically shaped specimens while cooling down to room temperature. A corresponding temperature/pressure profile is schematically displayed in **Figure 1**a (see methods for a detailed description of the high-temperature annealing experiment). We begin our discussion with one of the end cases, that is, an ErMnO$_3$ polycrystal annealed under the maximum pressure of $\sigma$ = 47 MPa reached in our experiment. The microstructure of this sample is analyzed via X-ray diffraction (XRD), electron backscattered diffraction (EBSD), and scanning electron microscopy (SEM) as summarized in Figure 1b. The XRD pattern reflects the hexagonal space group *P*6$_3$*cm* without any indication of secondary phases. The pole figure in the inset to Figure 1b is evaluated over ≈400 grains, showing a non-preferential crystallographic orientation of the grains.[29] The SEM micrograph outlines an isotropic grain shape, indicating the absence of mechanically induced high-temperature microstructural creep.

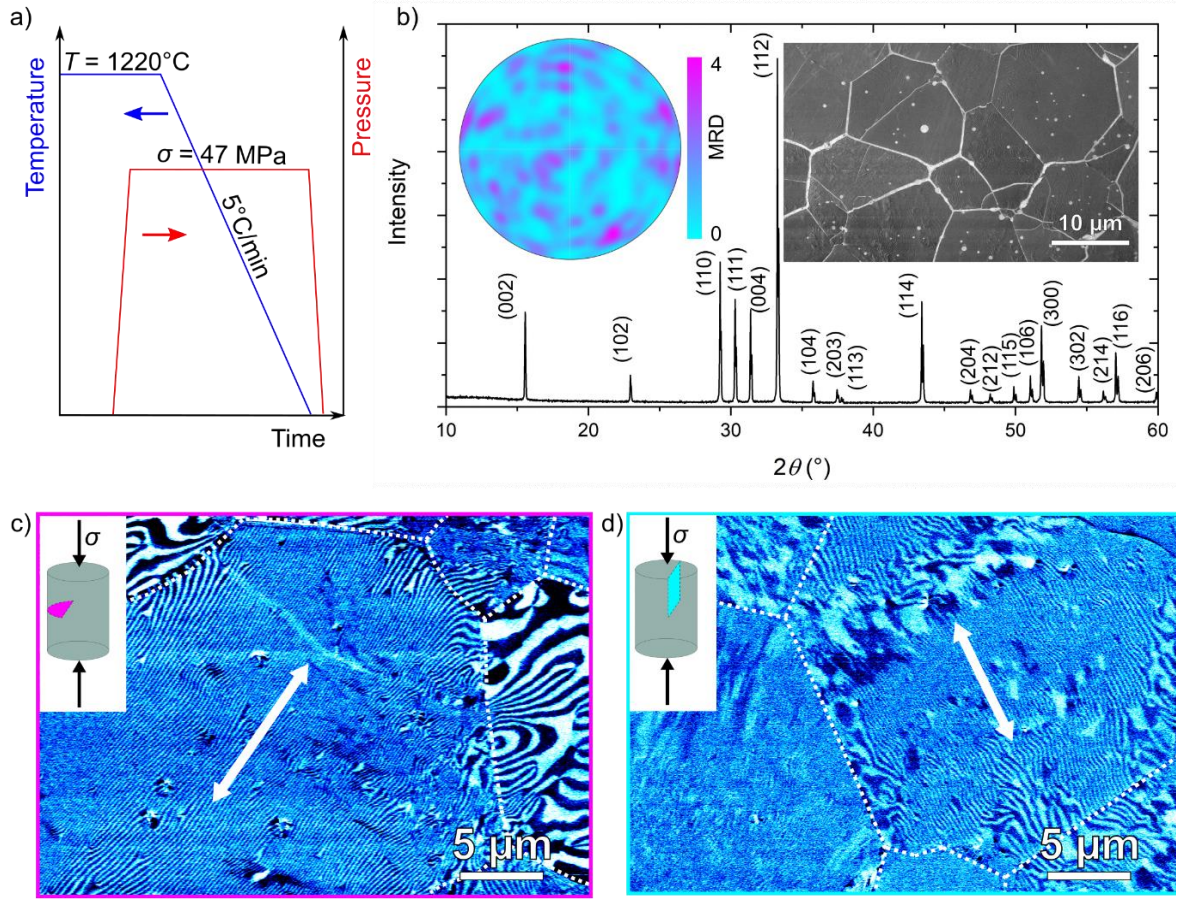

**Figure 1:** Ferroelectric domain structure in polycrystalline ErMnO$_3$ cooled under an applied mechanical pressure. a) Schematic temperature and pressure profile displaying the thermomechanical treatment of the sample from 1220°C to room temperature under an applied uniaxial pressure of $\sigma$ = 47 MPa. b) XRD pattern of ErMnO$_3$ after annealing. Characteristic (*hkl*) reflections are marked. The pole figure in the inset, obtained from EBSD orientation data from ≈400 grains, displays a uniform distribution of the {1000} planes with multiple of a random distribution (MRD) value < 4 for all orientations. A micrograph of the sample obtained from SEM is displayed in the inset. Lateral PFM contrast displaying the domain structure of a cross section c) perpendicular and d) parallel to the applied pressure. The grain boundaries are indicated as dotted white lines. The preferential orientation of the stripe-like domains within one grain along the white arrows are highlighted.

Next, we image the ferroelectric domain structure by piezoresponse force microscopy (PFM) with a peak-to-peak voltage of 10 V applied to the back of the sample at a frequency of 40.13 kHz. To access the 3D distribution of the domains, we investigate different cross-sections oriented perpendicular and parallel to the direction in which the mechanical pressure was applied during cooling. Corresponding images, showing the lateral PFM contrast, are displayed in Figure 1c and d, respectively. A pronounced PFM contrast is observed that allows for distinguishing the +*P* and −*P* domains (see, e.g., ref. [27] for technical details). For both cross-sections, we predominantly find periodic patterns of stripe-like domains with a periodicity of about 50 nm.

As highlighted by the white arrows, we consistently find a preferential orientation direction of the ferroelectric domain walls within one grain, which varies from grain to grain. Note that this behavior is completely different from ErMnO$_3$ single crystals[19, 30, 31] or polycrystalline samples cooled without a mechanical pressure,[27] where an isotropic vortex-like domain structure is predominant. The latter indicates a pronounced interaction between the domain formation and the applied uniaxial pressure, which we investigate systematically in the following.

## 2.2. Control of domain size and orientation

To understand the relation between applied mechanical pressure and the emergent domain structure, we perform annealing experiments under different mechanical pressures. To exclude pressure-induced changes of the crystal- and microstructure, we first record XRD patterns and SEM micrographs of the samples. We find that the crystal structure of all samples can be described by the space group $P6_3cm$, showing no indication of secondary phases (Figure S1). By analyzing about 20 grains in each sample, we measure an average grain size of 12.8±1.7 µm (Figure S2). Importantly, our data shows that the grain size is independent of the applied pressure, discarding mechanically induced microstructural creep and the resulting grain-size effects as the origin of variations in the domain structure.[32] Thus, going beyond previous grain-size dependent studies,[27] this sample series with grains of a single average size represents an ideal system for the investigation of pressure-driven phenomena.

Representative PFM images of the domain structure in samples cooled under different mechanical pressures are displayed in **Figure 2**a–c (larger overview images covering 50 x 50 µm$^2$ areas of parallel and perpendicular cross sections are displayed in Supplementary Figures S2). The domain structure displayed in Figure 2a ($\sigma$ = 0 MPa) is consistent with previous observations,[27] exhibiting a rather isotropic network of vortex domains within the different grains. The domain walls typically terminate at the grain boundaries and do not extend into adjacent grains, indicating that the ferroelectric domains in neighboring grains are largely independent as discussed in refs. [27] and [33]. Increasing the mechanical pressure during cooling to $\sigma$ = 24 MPa transforms the isotropic vortex-like domain structure into a stripe-like pattern (Figure 2b) in all grains, independent of their crystallographic orientation relative to the direction of the applied pressure. This stripe-like pattern reflects a preferred orientation for the domain walls within a single grain, while varying in direction from grain to grain. The response of the domain structure in ErMnO$_3$ to a mechanical pressure is unexpected, since the domain walls are purely ferroelectric. The qualitative change in the domain structure thus cannot be explained based on the release of mechanical pressure via the formation of ferroelastic domain walls as in conventional ferroelectrics, such as BaTiO$_3$, Pb(Zr,Ti)O$_3$ or (K,Na)NbO$_3$, pointing towards a different microscopic origin. As the comparison with PFM images acquired on samples annealed with a pressure of $\sigma$ = 47 MPa show (Figure 2c), the effect becomes even more pronounced as the mechanical pressure increases, leading to strongly elongated and highly ordered stripe-like domains within the different grains, as quantified by the pressure-dependence of the anisotropy parameter $A_\sigma$ obtained from the polar plots in Figure 2d. Note that the weak preferential orientation of the sample annealed under the absence of pressure is a consequence of the formation of stripe-like domains due to intergranular strain fields,[34] a known feature for polycrystalline ErMnO$_3$.[27] The transition to a stripe-like domain structure naturally leads to larger distances between the sixfold meeting points that form the characteristic vortex-/anti-vortex pairs in hexagonal manganites, highlighted by the dashed circles in Figure 2a–c.

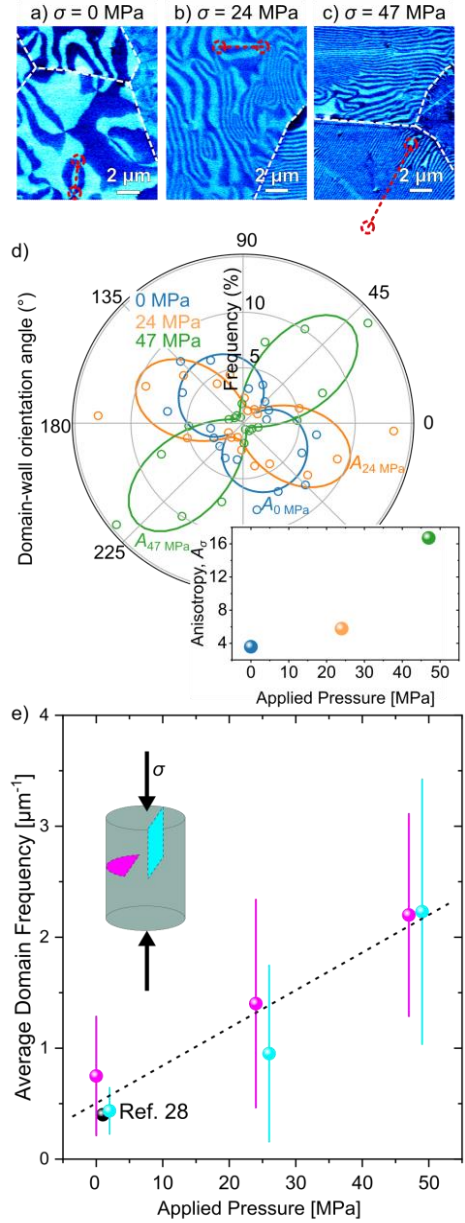

**Figure 2:** Control of the vortex density and domain frequency in polycrystalline materials via a mechanical pressure. Representative PFM images of polycrystalline ErMnO$_3$ cooled under a) 0 MPa, b) 24 MPa, and c) 47 MPa. Overview images over an area of 50 x 50 µm² of a section parallel and perpendicular to the mechanical pressure are displayed in Figure S3. Representative vortex-/anti-vortex pairs are marked by dashed red circles in a) and b), whereas the respective anti-vortex core is outside the displayed area in c). d) The relative distribution of the domain wall orientations is displayed for different mechanical pressures for a representative grain. The increasing degree of orientation of the stripe-like domains with increasing pressure is quantified by the anisotropy parameter $A_\sigma$, quantified in the inset (a derivation of $A_\sigma$ is provided in Figure S4). The average frequency of stripe-like domains is displayed as a function of the applied mechanical pressure in e). Cyan and purple data points represent the cross-section parallel and perpendicular to the applied mechanical pressure. A literature value of the average domain frequency of single crystalline counterparts cooled under the same cooling rate (5 K/min) is displayed for comparison.[28] The dashed line represents a linear fit to the experimental data.

To quantify the mechanically driven change in the ferroelectric domain structure, we evaluate next to their directional ordering (Figure 2d) the domain frequency. The domain frequency measures the number of domain intersections along the length of a test line drawn perpendicularly to the stripes, quantifying the size of the stripe-like domains.[35] In Figure 2e, we present the domain frequency for a cross-section parallel (cyan) and perpendicular (purple) to the applied pressure by averaging over 40 grains for each direction. The data indicates a one-to-one correlation between the applied mechanical pressure and the frequency of the stripe-like domain structure, showing an enhancement in frequency by a factor of ~4 as the mechanical pressure increases from 0 to 47 MPa. The same enhancement is observed in cross-sections oriented parallel (cyan) and perpendicular (purple) to the applied pressure. The latter leads us to the conclusion that the orientation of the emergent stripe domains and associated domain walls is determined by the crystallographic orientation of the different grains rather than the direction of the applied uniaxial pressure, which we will elaborate on in the following.

A more detailed analysis of the crystallographic orientation of the domains of our polycrystalline ErMnO$_3$ (Figure 1c) in terms of spatially correlated EBSD and PFM measurements is presented in **Figure 3**. EBSD measurements of a selected area are displayed in Figure 3a, where the orientation of the grains is indicated schematically by hexagonal prisms. A PFM image (in-plane contrast) of the area marked by the dashed black box in Figure 3a is given in Figure 3b. The data suggests that the stripe-like domains and associated domain walls predominantly align parallel to the crystallographic $c$-axis ($P \parallel c$). To corroborate this correlation, we analyze 29 grains and plot the domain wall orientation obtained from the PFM scans against the orientation of the $c$-axis measured by EBSD in Figure 3c (orientations are measured with respect to the same reference plane as explained in Figure S5). Figure 3c confirms that there is strong correlation between the pressure-induced stripe-like domain pattern and the crystallographic orientation of the individual grains, revealing a preference to align the walls parallel to the polar $c$-axis when pressure is applied during annealing.

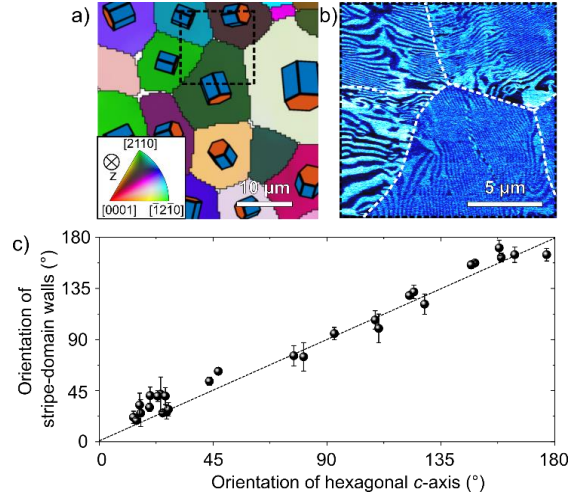

**Figure 3:** Control of the orientation of stripe-like domains. a) EBSD image of polycrystalline ErMnO$_3$ thermomechanically treated under a pressure of 47 MPa (cross section perpendicular to the mechanical pressure) displays the individual grains with their respective 3D direction as indicated schematically by the hexagonal prisms. An in-plane PFM image of the position marked with the black square is displayed in b). The grain boundaries are highlighted as dashed white lines. Comparing the EBSD and PFM images indicates that the orientation of the stripe-like domains is parallel to the $c$-axis of the hexagonal prism. The correlation between the orientation of the stripe-like domains and the $c$-axis of the hexagonal prism is displayed in c), systematically analyzed for 29 grains with preferable in-plane orientation. An explanation of the quantification of the data for one representative grain is sketched in Figure S5. The dashed black line indicates the ideal parallel orientation between the ferroelectric domain wall and the $c$-axis of the prism.

## 2.3. Domain structure

To understand the pressure-induced changes in the domain structure and their relation to the crystallographic structure, we perform phase-field simulations. Following the established phase-field model for hexagonal manganites,[36, 37] we represent the system by two parameters, i.e., the trimerization amplitude $Q$ and phase $\phi$ as elaborated in the method section. The model reproduces the characteristic vortex-like domain structure of ErMnO$_3$ as displayed in **Figure 4**a, where the ferroelectric polarization is parallel/antiparallel to the $c$-direction. The phase-field simulation allows for investigating the impact of uniaxial pressure within the volume, giving a 3D model of the pressure-induced domain structure. For this purpose, we consider the coupling between the energy density and the mechanical pressure,[36]

$$f_{\text{strain}} = GQ^2[(u_{xx} - u_{yy})\delta_x\Phi - 2u_{xy}\delta_y\Phi] \quad (\text{eq. 1})$$

where $G$ is the strain coupling coefficient and $(x, y)$ are the Cartesian coordinates in the $xy$ plane. Within the applied model (eq. 1), the strain does not couple directly to the domain structure, which is fundamentally different from ferroelastic ferroelectrics.[38] Instead, as described by eq. 1, the elastic strain creates a force that pulls the vortices and anti-vortices away from each other.[26, 37] The difference between the strain components, $u_{xx} - u_{yy}$, results in a modulation of the domain structure in the $x$-direction, whereas the shear strain component, $u_{xy}$, results in a modulation in the $y$-direction. According to eq. 1, the effect depends on the magnitude of the strain components and the respective phase gradients, $\delta_x \Phi$ and $\delta_y \Phi$.

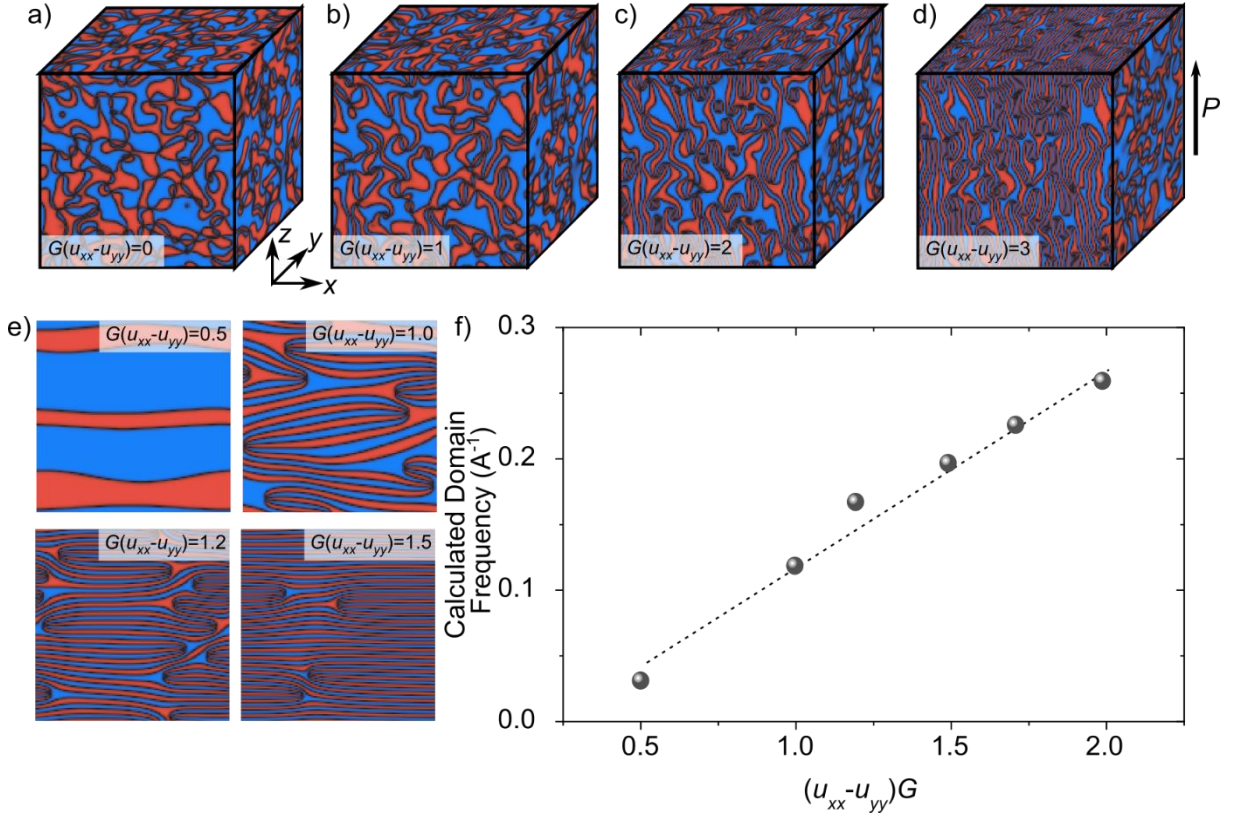

**Figure 4:** Phase-field simulations showing the relation between mechanical pressure and the domain morphology in ErMnO$_3$. The influence of the magnitude of applied pressure, $G(u_{xx} - u_{yy})$, on the 3D domain structure is displayed in a) to d). The ferroelectric polarization, $P$, is parallel/antiparallel to the $z$-direction. The preferential formation of a layer-like domain structure with increasing magnitude of the pressure can be observed. The correlation between the applied pressure and the domain periodicity is displayed in e) and quantified in f). The dashed line represents a linear fit to the simulated data.

Our experimental parameter, that is, the mechanical pressure, is linked to the distribution of the individual strain components in the phase-field simulations, $u_{xx}$, $u_{yy}$, and $u_{xy}$, via the anisotropic Youngs Modulus.[39] The 3D domain structure simulated for varying pressure amplitudes is displayed in Figure 4c–d. In agreement with our experimental findings (Figure 2a–c), the simulations reproduce a pressure-driven transition from an isotropic vortex-like (Figure 4a) to an anisotropic stripe-like (Figure 4d) domain structure. In comparison to previous two-dimensional simulations on ErMnO$_3$,[27, 36, 37] our three-dimensional phase-field model highlights that the stripe-like ferroelectric domains extend into an oriented layer-like structure in the third dimension. Consistent with the experimental data in Figure 3c, the simulations show that the pressure-induced stripe-like domains and associated walls preferentially orient parallel to the polar axis of the system. As expressed by eq. 1, this is a consequence of the absence of the coupling of the energy density to the strain in $z$-direction, i.e., $u_{zz}$. Physically, strains along the polar axis are invariant under all symmetry operations of the ErMnO$_3$ system and, hence, merely result in a small shift of the Landau expansion parameters. Hence, these terms are usually ignored for Landau expansions in any order. [36] While we can explain the experimentally observed emerging domain structures based on eq. 1, additional electrostatic effects[40] at the domain walls may further promote the formation of the stripe-like domain structure, as the associated walls are predominantly electrically neutral. Importantly, based on the phase-field simulations, we can correlate the thickness of each stripe-like domain to the strain-magnitude. 2D representations of the domain structure are displayed in Figure 4e and the simulated domain frequency, $f$, is displayed as a function of the applied pressure, $G(u_{xx} - u_{yy})$, in Figure 4f. We find that the domain frequency continuously increases with the applied pressure, which is consistent with the experimentally observed relationship (Figure 2e). Thus, the mechanical pressure provides a lever to transfer isotropic vortex-like domains into a periodic layered domain pattern with tunable orientation and domain size.

## 3. Conclusion

Engineering the domain structure of ferroelectric materials is crucial in order to enhance their dielectric and piezoelectric properties and the backbone of their application as capacitors, sensors, and in energy storage devices.[41] Conventionally, domain engineering is realized via the crystal- or microstructure[1] and chemical composition[42] of the material, or controlled via elastic strain.[12] Thus, most existing concepts rely on the interaction between elastic strains and the ferroelastic domains, which ultimately limits them to materials exhibiting ferroelasticity. Unlocked by elastic-strain induced forces acting on the vortex-/anti-vortex pairs,[36] domain engineering was recently realized utilizing elastic strain fields originating from confinement effects in non-ferroelastic ErMnO$_3$.[27] Here, we expand this concept towards pressure-driven domain engineering, providing a conceptually new lever beyond microstructural effects. We find a one-to-one correlation between the applied pressure and the frequency and orientation of the induced stripe-like domains. Most importantly, and different from ferroelastic ferroelectrics, such as BaTiO$_3$, Pb(Zr,Ti)O$_3$ or (K,Na)NbO$_3$, our pressure-engineering approach does not require ferroelasticity. On the one hand, removing the need for ferroelastic domain walls expands the pool of candidate materials. On the other hand, it foreshadows a conceptually different way for mechanical switching of ferroelectrics of improper nature that does not result in unwanted changes in the physical shape of the sample due to the movement of ferroelastic domain walls.[43] This would be beneficial for the lifetime of the ferroelectric material and crucial for its application in memory devices or as multilayer ceramic capacitors.

## 4. Method Section

*Solid state synthesis:* Synthesis of ErMnO$_3$ powder was done by a solid-state reaction of dried Er$_2$O$_3$ (99.9% purity; Alfa Aesar, Haverhill, MA, USA) and Mn$_2$O$_3$ (99.0% purity; Sigma-Aldrich, St. Louis, MO, USA). The powders were mixed and ball milled (BML 5 witeg Labortechnik GmbH, Wertheim, Germany) for 12 hrs at 205 rpm using yttria stabilized zirconia milling balls of 5 mm and ethanol as dispersion medium. The reaction to ErMnO$_3$ was done by stepwise heating at 1000°C, 1050°C, and 1100°C for 12 hrs. More details on the powder processing can be found in ref. [27]. The powder was isostatically pressed into samples of cylindrical shape at a pressure of 200 MPa (Autoclave Engineers, Parker-Hannifin, Cleveland, OH, USA). Sintering was carried out in a closed alumina crucible with sacrificial powders of the same chemical compositions at a temperature of 1400°C for 4 hrs with a heating and cooling rate of 5°C/min (Entech Energiteknik AB; Ängelholm, Sweden).

*Thermomechanical treatment:* The cylinders with a diameter of ~7.6 mm and a height of ~5.1 mm were put into a Al$_2$O$_3$ die filled with coarse MgO powder to prevent the sample reacting with the die during heat treatment. The samples were heated to a temperature of 1220°C, thus above $T_c$ of ErMnO$_3$ ($T_c$~1156°C [28]) with a heating rate of 5°C/min. After a dwell time of 30 minutes to thermally equilibrate the sample, a uniaxial pressure of either 0 MPa, 24 MPa, and 47 MPa was applied to the die, while maintaining a temperature of 1220°C for 10 more minutes. As illustrated in Figure 1a, the samples were cooled with the applied pressure until room temperature. For all experiments, a constant cooling rate of 5°C/min was utilized.

*Structural and microstructural characterization:* After thermomechanical treatment, the samples were cut into half-discs using a diamond wire saw (Serie 3000, Well, Mannheim, Germany) to get insight into the micro- and nanostructure parallel and perpendicular to the applied mechanical pressure. Next, the samples were lapped with a 9 µm-grained Al$_2$O$_3$ water suspension (Logitech Ltd., Glasgow, UK) and polished using silica slurry (Ultra-Sol® 2EX, Eminess Technologies, Scottsdale, AZ, USA). The crystal structure and phase purity of our samples were determined by XRD (Bruker, Billerica, MA, USA). Micrographs of the polished samples were acquired by performing SEM (Helios G4 UX, FEI, Lausanne, Switzerland). The nanoscale domain structure of the samples was obtained via PFM (NT-MDT, Moscow, Russia) with an electrically conductive tip (Spark 150Pt, NuNano, Bristol, UK). The samples were excited using an alternating voltage with a frequency of 40.13 kHz and an amplitude of 10 V. The laser deflection was read out by two Lock-in amplifiers (SR830, Stanford Research Systems, Sunnyvale, CA, USA). Prior to PFM measurement on the polycrystalline ErMnO$_3$, the PFM response was calibrated on a periodically out-of-plane poled LiNbO$_3$ sample (PFM03, NT-MDT, Moscow, Russia).

To map the orientation of the grains, EBSD was performed (Ultra 55 FEG-SEM, Zeiss, Jena, Germany). To obtain sufficient statistics an area of 400 grains (approximately 235x235µm²) was mapped, whose domain structure was characterized by PFM prior to the EBSD analysis. Before the EBSD scan, the sample was carbon coated. The scan was performed with an acceleration voltage of 10 kV, a working distance of 20.4 mm and a sample rotation of 70°. Kikuchi diffraction patterns of 120x120 px² were recorded with a nominal step size of 0.5 µm. Diffraction patterns were indexed by dictionary indexing[44] followed by refinement as implemented in kikuchipy (version 0.8.0),[45] based on dynamical simulations of ErMnO$_3$ from EMsoft (version 5)[46]. Orientations used in dictionary indexing were sampled from the Rodrigues Fundamental Zone of proper point group 622.[47] Orientation analysis was performed in Matlab with MTEX (version 5.8).[48]

*Phase-field simulations.* We perform phase-field simulations based on the Landau expansion of the free energy $F(Q, \Phi, P, u_{xx}, u_{yy}, u_{xy})$. The Landau expansion is obtained by adding the ferroelectric and the strain terms of the Landau expansion.[36, 49] We choose the parameters of the ferroelectric term as in ref. [36], with the exception of $s_P^x = 8.88$ eV to ensure stability. The parameter $G$ of the strain term is set as described in Figure 4 in the main text. For all simulations, we choose a uniform Cartesian, three-dimensional grid with lattice spacing $d_x = d_y = 0.2$ nm and $d_z = 0.3$ nm as computational mesh. We use periodic boundary conditions in the *x* and *y*-directions and open boundary conditions in the *z*-direction. We fix a uniform strain throughout the entire system and derive the Ginzburg-Landau equations for the structural and ferroelectric order parameters.[50] The Ginzburg-Landau equations are then integrated in a finite difference integration scheme using a Runge-Kutta 4 integrator with time step $\Delta t = 5 \cdot 10^{-4}$. The simulations of the three-dimensional crystals have been performed with a mesh of size $n_x \times n_y \times n_z = 128 \times 128 \times 128$. An initial, randomly generated field of all order parameters is evolved with uniform zero strain for $6 \cdot 10^{-4}$ time steps, such that a domain structure is formed. Then, the system is further integrated under uniform, constant strain for a time of $1.2 \cdot 10^5$ time steps.

To determine the stripe domain size, simulations with a mesh of size $n_x \times n_y \times n_z = 128 \times 128 \times 2$ are performed. Again, an initial, randomly generated field of all order parameters is evolved with uniform zero strain for $1.6 \cdot 10^5$ time steps to form an initial domain pattern. Then, the systems are integrated with uniform strain until the domains form a stripe-like pattern and no longer evolve. The integration time varies with the applied strain. Systems with lower strain evolve more slowly and therefore require more time steps. Finally, the stripe domain size is extracted using a stereological method.[35] Note that the calculated domain frequency (Figure 4f) is smaller in comparison to the experimentally determined values (Figure 2e), which we relate to the fact that we set the parameter $G = 1$ in equation 1 for simplicity.

## Acknowledgements

J. Cilensek and A. Debevec are kindly acknowledged for performing the experiments using the hot press. S. Artyukhin and G. Catalan are acknowledged for helpful discussions. J.S. acknowledges the support of the Alexander von Humboldt Foundation through the Feodor-Lynen fellowship. J.S. and D.M. acknowledge NTNU Nano for the support through the NTNU Nano Impact fund. D.M. thanks the NTNU for support through the Onsager Fellowship Program, the outstanding Academic Fellow Program, and acknowledges funding from the European Research Council (ERC) under the European Union's Horizon 2020 Research and Innovation Program (Grant Agreement No. 863691). H.W.Å. acknowledges NTNU for financial support through the NTNU Aluminum Product Innovation center (NAPIC). M.Z. acknowledges funding from the Studienstiftung des Deutschen Volkes via a Doctoral Grant and the State of Bavaria via a Marianne-Plehn scholarship. T.R. would like to acknowledge the Slovenian Research Agency (research core funding P2-0105).

**Conflict of Interest**

The authors declare no conflict of interest.

Supporting Information

# Pressure-control of non-ferroelastic ferroelectric domains in ErMnO$_3$

O. W. Sandvik[1], A. M. Müller[2], H.W. Ånes[1], M. Zahn[1,3], J. He[1], M. Fiebig[2], Th. Lottermoser[2], T. Rojac[4], D. Meier[1,*], and J. Schultheiß[1,3,*]

[1] Department of Materials Science and Engineering, Norwegian University of Science and Technology (NTNU), 7034 Trondheim, Norway

[2] Department of Materials, ETH Zurich, 8092 Zurich, Switzerland

[3] Experimental Physics V, University of Augsburg, 86159 Augsburg, Germany

[4] Electronic Ceramics Department, Jožef Stefan Institute, 1000 Ljubljana, Slovenia

*corresponding authors: dennis.meier@ntnu.no; jan.schultheiss@uni-a.de

## 1. Structural and microstructural analysis

### 1.1. Crystallography

X-Ray diffraction (XRD) patterns of polycrystalline ErMnO$_3$ cooled under different mechanical pressures are displayed in Figure S1. The crystal structure of all samples can be described with the space group *P*6$_3$*cm*, while secondary phases are absent.

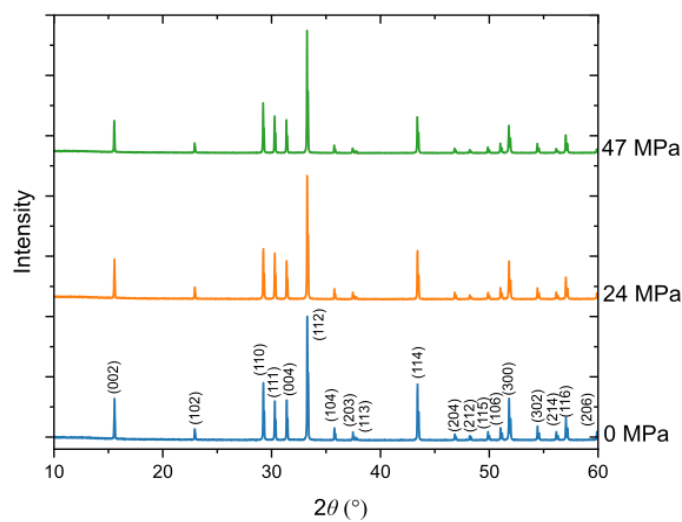

**Figure S1:** XRD pattern of polycrystalline ErMnO$_3$ cooled under different mechanical pressures.

## 1.2. Microstructure

Scanning electron microscopy (SEM) micrographs of polycrystalline ErMnO$_3$ cooled under different mechanical pressures are displayed in Figure S2. A systematic grain size analysis over 20 grains in each sample finds that the grain size is independent of the applied mechanical pressure and an average grain size of $g=12.8\pm1.7$ μm was identified ($g_{0\text{ MPa}} = 13.3 \pm 1.4$ μm, $g_{24\text{ MPa}} = 10.9 \pm 1.4$ μm, and $g_{47\text{ MPa}} = 14.2 \pm 0.6$ μm). The microstructures features a mixture of inter- and intragranular microcracks, commonly found in polycrystalline hexagonal manganites, originating from a combination of a large volume change and a strong anisotropy in the thermal expansion coefficient of the hexagonal structure.[1, 2] In addition, ferroelectric domain and domain wall contrast can be observed in the SEM images, originating from electron emission yield as explained in ref. [3].

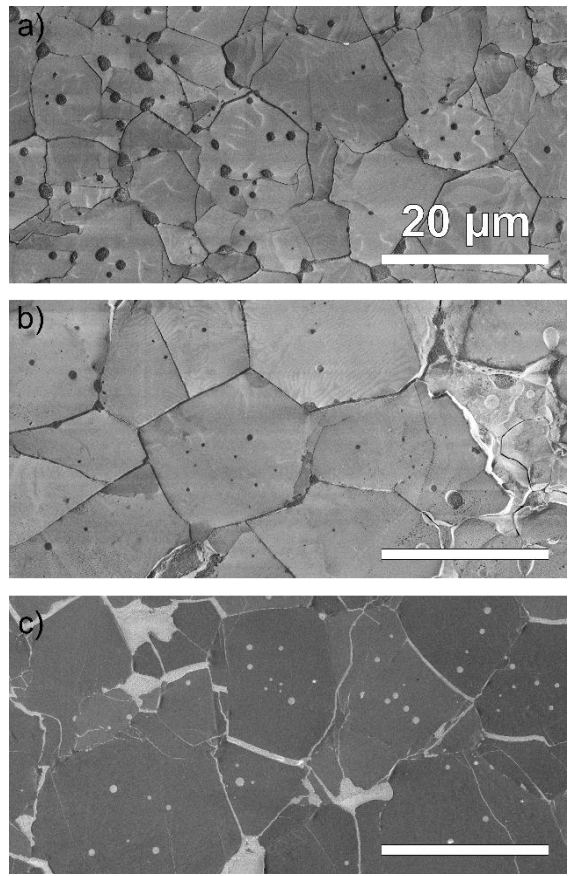

**Figure S2:** SEM images of polycrystalline ErMnO$_3$ cooled under different mechanical pressures. a) 0 MPa, b) 24 MPa, and c) 47 MPa.

## 2. Domain structure analysis via PFM

Piezoresponse Force Micrsocopy (PFM) images of a ferroelectric domain structure in polycrystalline ErMnO$_3$ cooled under different mechanical pressure, measured over an area of 50 x 50 µm² are displayed in Figure S3.

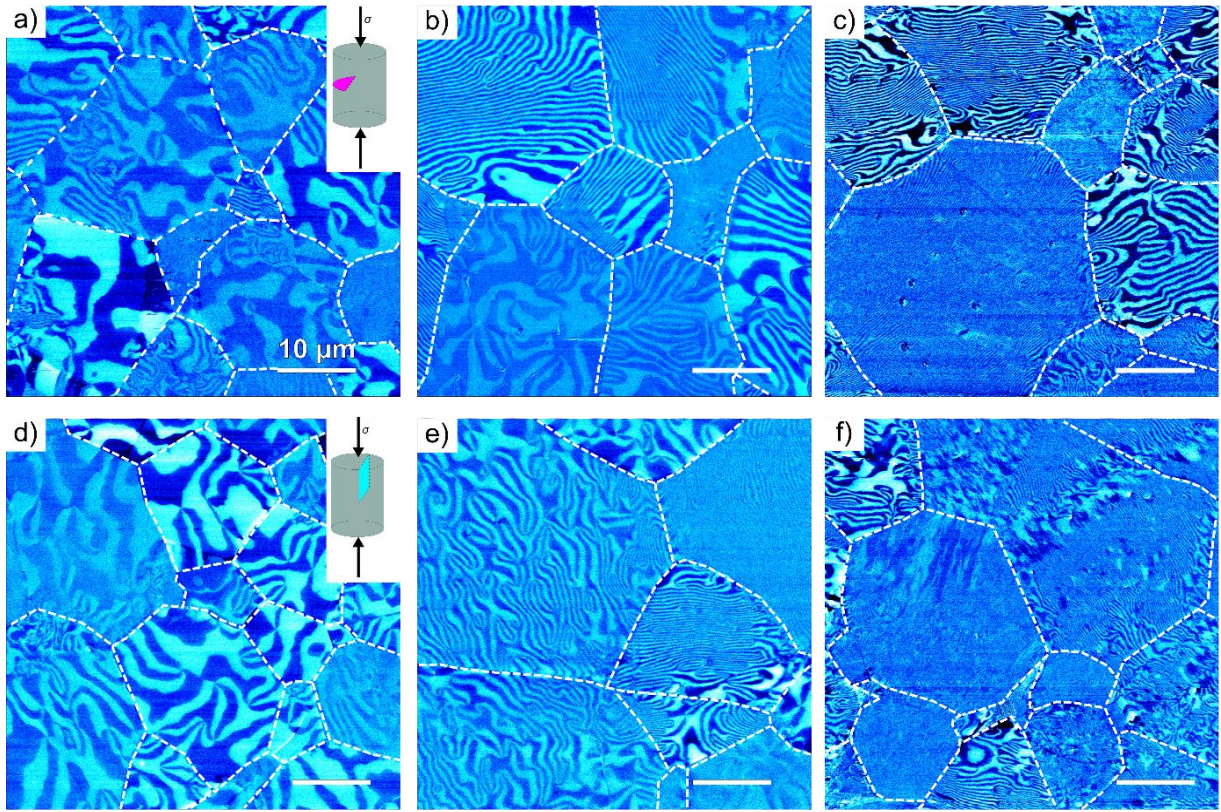

**Figure S3:** PFM images of polycrystalline ErMnO$_3$ samples cooled under different mechanical pressures. Images obtained from perpendicular and parallel cross sections are displayed in a)-c) and d)-f), respectively. The applied mechanical pressure is $\sigma = 0$ MPa (a) and d)), $\sigma = 24$ MPa (b) and e)), and $\sigma = 47$ MPa (c) and f)).

The strategy to extract the domain wall orientations from individual grains for the PFM images is explained in Figures S4.

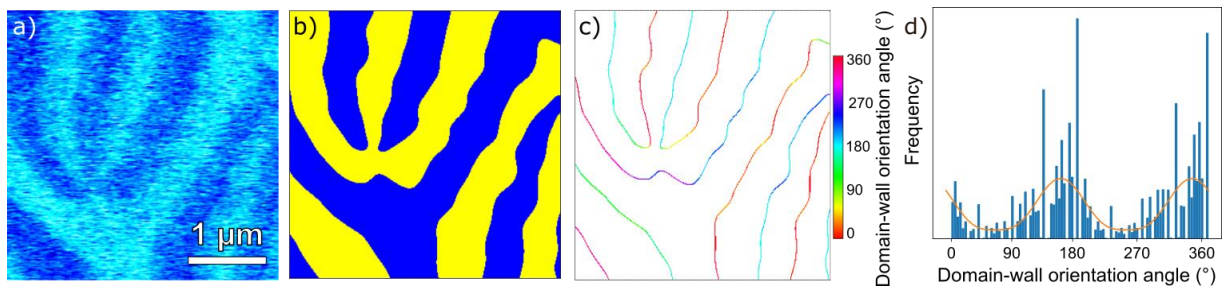

**Figure S4:** Extraction of domain wall orientations from PFM images, for an example of the sample annealed at $\sigma=24$ MPa. a) Section the PFM image. As first a Gaussian filter is applied

to reduce random noise. b) Via local thresholding using Otsu's method,[4] a binary image is obtained. Further, small islands of both colors with sizes much smaller than the characteristic length scale of the domains are removed. c) By applying Canny edge detection[5] and taking the local 2D derivative of the image, the position and local in-plane orientation of each domain wall segment is determined. d) Histogram of the in-plane domain wall orientation angles. Two peaks in opposite directions indicate the preferred domain wall orientation. The probability distribution is fitted to $A_\sigma \cdot (\cos^2(\phi - \phi_0))^w + 1$, where $A_\sigma$ characterizes the height of the peaks compared to the flat regions, $w$ the width of the peaks and $\phi_0$ the preferred wall orientation. The probability distribution is inspired by Malus' law that describes the transmittance of a polarizer and fulfills similar properties of symmetry.

To understand the origin for the preferential orientation of the stripe-like domain walls, PFM, and electron backscattered diffraction (EBSD) data are collected on the same position. The orientation angle of the hexagonal c-axis, $\alpha_{EBSD}$, is quantified from EBSD data, while the orientation angle of the stripe domain walls, $\alpha_{PFM}$, is obtained from PFM measurements. The quantification of these angles from the raw data is displayed for one grain in Figure S5.

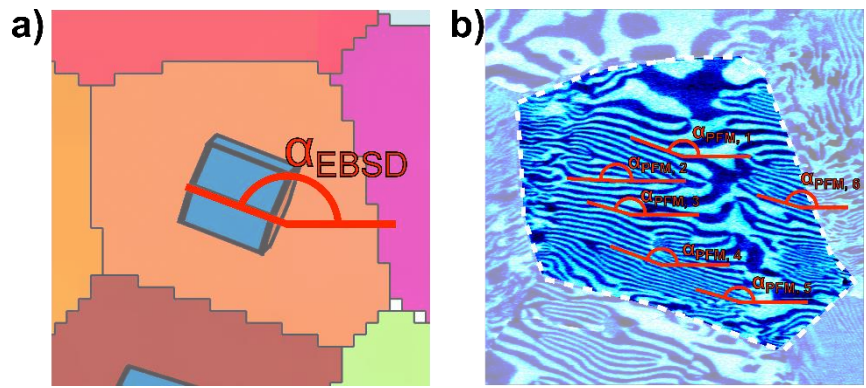

**Figure S5:** Quantification of the orientation angle of the a) hexagonal *c*-axis from EBSD data $\alpha_{EBSD}$, and the b) orientation of the stripe domain walls from PFM data, $\alpha_{PFM}$, sketched for one representative grain. To determine the orientation of the stripe domain walls, an averaged value from several domain walls is obtained.